\documentclass[pre,twocolumn,showpacs,showkeys]{revtex4}
\usepackage{epsfig,epsf,graphics,color,rotate,amsmath}

\begin{document}

\title{Bound of dissipation on a plane Couette dynamo}
\date{\today}
\author{Thierry \surname{Alboussi\`ere}} 
\affiliation{ Laboratoire de G\'eophysique Interne et Tectonophysique, CNRS, Observatoire de Grenoble, Universit\'e Joseph Fourier, Maison des G\'eosciences, BP 53, 38041 Grenoble Cedex 9, France}
\email{thierry.alboussiere@ujf-grenoble.fr}

\begin{abstract}
Variational turbulence is among the few approaches providing rigorous results in turbulence. In addition, it addresses a question of direct practical interest, namely the rate of energy dissipation. Unfortunately, only an upper bound is obtained as a larger functional space than the space of solutions to the Navier-Stokes equations is searched. Yet, in some cases, this upper bound is in good agreement with experimental results in terms of order of magnitude and power law of the imposed Reynolds number. In this paper, the variational approach to turbulence is extended to the case of dynamo action and an upper bound is obtained for the global dissipation rate (viscous and Ohmic). A simple plane Couette flow is investigated. For low magnetic Prandtl number $P_m$ fluids, the upper bound of energy dissipation is that of classical turbulence ({\it i.e.} proportional to the cubic power of the shear velocity) for magnetic Reynolds numbers below $P_m^{-1}$ and follows a steeper evolution for magnetic Reynolds numbers above $P_m^{-1}$ ({\it i.e.} proportional to the shear velocity to the power four) in the case of electrically insulating walls. However, the effect of wall conductance is crucial: for a given value of wall conductance, there is a value for the magnetic Reynolds number above which energy dissipation cannot be bounded. This limiting magnetic Reynolds number is inversely proportional to the square root of the conductance of the wall. Implications in terms of energy dissipation in experimental and natural dynamos are discussed.
\end{abstract}

\maketitle

\section{Introduction}

Natural dynamos exist whenever the conditions of their existence is possible, {\it i.e.} when a sufficiently large magnetic Reynolds is reached. One can imagine that this is an additional route for mechanical energy dissipation and that it is more likely for natural systems to take it than not to take it. 
It has been argued sometimes that a state of maximal dissipation rate should be reached and this idea has been used as a closure assumption for turbulence ({\it e.g.} Malkus \cite{malkus03}). It is not necessary to make such an assumption,  as explained clearly by Howard \cite{howard}, and yet one can draw useful information from the determination of rigorous lower and upper bounds on energy dissipation in turbulent flows. These bounds are obtained in a larger functional space than the solutions to the Navier-Stokes equation, hence they are not necessarily attained. However, in a number of cases, turbulent flows lead to a dissipation rate of the same magnitude as this upper bound \cite{PK03}.  

The variational approach to turbulence was introduced by Malkus \cite{malkus54}, Busse \cite{busse69} and Howard \cite{howard63}. More recently, this approach was reformulated by Doering and Constantin \cite{DC92} and expressed in a simpler way, using the concept of a background function (not necessarily the mean flow of turbulence) following Hopf \cite{hopf41}. The objective is to bound energy dissipation under the constraint of horizontally averaged energy balance (for statistically plane invariant configurations). In a series of papers, the approach was improved by optimizing the spectral Lagrange parameter \cite{NGH97} and the background function \cite{PK03}. The final bound is better than the nearly rigorous bound of Busse. 
Our objective here is not to exhaust these possibilities of optimization but rather to apply the general principle to a new configuration, {\it i.e.}  a dynamo problem. 

The idea of applying a variational approach to a magnetohydrodynamic flow has been applied already to a Couette and Poiseuille flow subjected to an applied transverse magnetic field \cite{APMD03}. This paper has been a source of motivation for the present work with two significant variations. First, there is now no imposed magnetic field and secondly, magnetic boundary conditions are different. In the paper by Alexakis {\it et al.}, magnetic disturbances are constrained to vanish at the boundaries, which does not correspond to a physically plausible situation. It is assumed here that the fluid domain is bounded by an infinite domain of electrically insulating medium (or with a conducting solid wall of finite thickness in between).


A plane Couette flow configuration is considered and our objective is to find an upper bound to the total energy dissipation when a prescribed velocity is applied. There is no applied magnetic field of external origin and the problem may look like a purely hydrodynamical one. However, the flow may support dynamo action (see \cite{WB02,RK03}) and the amount of dissipated energy must then take into account Joule dissipation. Upper bounds will be obtained as a function of two dimensionless parameters, the Reynolds number and the magnetic Reynolds number. In addition, the effect of electrically conducting walls of finite thickness will be investigated. 

Section \ref{configuration} provides details on the flow configuration, notation, dimensionless variables and equations. In section \ref{background}, the principle of decomposition with background flow is presented. The horizontally averaged energy balance is obtained in section \ref{ebalance} and the expression for the total energy dissipation (Joule plus viscous) is given in section \ref{totenergy}. Energy dissipation bounds are obtained in section \ref{velspec} and \ref{magspec} respectively, when velocity fluctuations and magnetic fluctuations are considered respectively in addition to the background flow. An improved bound is determined numerically in section \ref{matlabopt}.
Section \ref{discussion} is devoted to a discussion of the bounds obtained and their relevance to experimental and geophysical configurations and section \ref{perspectives} to the directions in which the variational approach could be extended to deal with more relevant models of dynamo. 


It may be useful to provide some guidance on how to read this paper. 
The Hopf-Doering-Constantin method is explicitly introduced and subsequent calculations of upper bounds are also detailed explicitly. A reader with no prior knowledge of the method can check all results with pen and paper until the end of section \ref{magspec}. Section \ref{matlabopt} does not provide any fundamentally new result and can be ignored in a first reading. This section has required some standard numerical calculations of eigenvalues related to the magnetic spectral constraint, and the method is only sketched. This section serves two purposes: first, it provides a better upper bound as it corresponds to a background function for which the spectral condition is only just satisfied (it is zero for a particular disturbance) and secondly, it provides a confirmation that the upper bound derived analytically in section \ref{magspec} is relevant as it obeys the same scaling law as the numerical bound at large magnetic Reynolds numbers.  

\section{Plane Couette flow configuration}
\label{configuration}

\begin{figure}
\begin{center}
\includegraphics[width=8.5 cm, keepaspectratio]{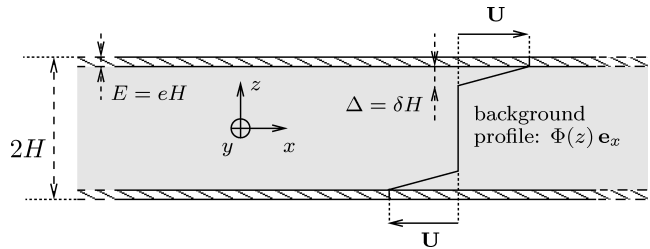}
\caption{Plane Couette flow configuration}
\label{configurationfig}
\end{center}
\end{figure}

The dimensionless Navier-Stokes and induction equations can be written
\begin{eqnarray}
\frac{\partial {\bf u}}{\partial t} + {\bf u}\cdot {\bf \nabla} {\bf u} = -{\bf \nabla } p + {\bf j} \times {\bf B} + Re^{-1} {\bf \nabla }^2 {\bf u}, \label{ns} \\
\frac{\partial {\bf B}}{\partial t} + {\bf u}\cdot {\bf \nabla} {\bf B} = {\bf B}\cdot {\bf \nabla} {\bf u} + R_m^{-1}  {\bf \nabla }^2 {\bf B}. \label{induction}
\end{eqnarray}
In the equations above, the dimensional length and velocity scales are chosen to be half the distance, $H$, and differential velocity, $U$, between the plates (see Fig. \ref{configurationfig}) while the quantities $H/U$, $\rho U^2$, $U \sqrt{\rho \mu}$, $\sqrt{\rho / \mu} U/H$ are taken as dimensional scales for time, pressure, magnetic field and electric current density respectively. The dimensionless parameters are the Reynolds number $Re = UH/\nu$ and the magnetic Reynolds number $R_m = \mu \sigma U H $. The symbols $\rho$, $\mu$, $\nu$ and $\sigma$ denote density, vacuum magnetic permeability, kinematic viscosity and electrical conductivity respectively.  
In this dimensionless formulation, the electric  current density ${\bf j} $ is simply the curl of the magnetic field ${\bf B}$. 

As shown on Fig. \ref{configurationfig}, the thickness of the fluid layer is $2H - 2 E$ where $E$ is the thickness of the electrically conducting layer of the wall. That thickness may be zero $E=0$, in which case the wall is electrically insulating. In all cases, the overall thickness of the electrically conducting domain is $2H$. For simplicity, the electrical conductivity of the solid conducting layer is identical to that of the fluid. Hence, at $z=H-E$ and $z=-H+E$, there is continuity of all three components of the magnetic field. Moreover, the induction equation is identical in the fluid and in the conducting layer. Therefore, there is no boundary condition to consider at $z=H-E$ or $z=-H+E$. The magnetic boundary condition, at $z=\pm H$, is that the magnetic field in the electrically conducting domain is matched continuously to a curl-free and divergence-free magnetic field outside (see appendix \ref{appendix2}).     

\section{Background velocity decomposition}
\label{background}

A frame of reference $(x,y,z)$ is chosen with $x$ aligned with the direction of the imposed shear velocity and $z$ in the direction perpendicular to the plates. Its origin is half-way between the plates.\\

The velocity field ${\bf u}$ is written as the sum of a steady parallel profile $\Phi (z) {\bf e}_x$ satisfying the boundary conditions $\Phi (\pm (1 - e)) = \pm 1$ and of a field ${\bf v}$ with homogeneous boundary conditions ${\bf v}(x,y,\pm (1 - e),t) = {\bf 0}$, {\it i.e.}
\begin{equation}
{\bf u} (x,y,z,t) = \Phi (z) {\bf e}_x + {\bf v} (x,y,z,t)  \label{decomposition}
\end{equation}
It must be stressed here that $\Phi$ -- the so-called background velocity profile -- needs not be the average value of the full velocity field ${\bf u}$, neither in the temporal nor in the horizontal space average sense. \\

Substituting this decomposition into the Navier-Stokes and induction equations leads to evolution equations for ${\bf v}$ and ${\bf B}$ for any particular choice of a background profile:
\begin{eqnarray}
&&\frac{\partial {\bf v}}{\partial t} + {\bf v}\cdot {\bf \nabla} {\bf v} + \Phi \frac{\partial {\bf v}}{\partial x} + v_z \Phi ' {\bf e}_x = -{\bf \nabla } p + {\bf j} \times {\bf B} \hspace*{10 mm} \nonumber \\
&& \hspace*{40 mm} + Re^{-1} \left[ \Phi '' {\bf e}_x + {\bf \nabla }^2 {\bf v} \right], \label{fluctv} \\
&&\frac{\partial {\bf B}}{\partial t} + {\bf v}\cdot {\bf \nabla} {\bf B} + \Phi \frac{\partial {\bf B}}{\partial x} = {\bf B}\cdot {\bf \nabla} {\bf v} + B_z \Phi ' {\bf e}_x \hspace*{15 mm} \nonumber \\
&& \hspace*{58 mm} + R_m^{-1}  {\bf \nabla }^2 {\bf B}, \label{fluctB}
\end{eqnarray}
where $\Phi '$ and $\Phi ''$ denote the first and second derivative of the background velocity profile $\Phi$ respectively. \\

\section{Energy balance of fluctuations}
\label{ebalance}

The dot product of equations (\ref{fluctv}) and (\ref{fluctB}) with ${\bf v}$ and ${\bf B}$ respectively are integrated over ${\cal{V}}$ a large rectangular volume $-L < x < L$, $-L< y < L$ and $-1 < z < 1$. They are then integrated over a long period of time $T$. The following space and time averaged equations are obtained:
\begin{eqnarray}
 \left< \Phi ' v_x v_z \right>  = \left< ( {\bf j} \times {\bf B} ) \cdot {\bf v} \right> - Re^{-1} \left<  \Phi ' \frac{ \partial v_x}{\partial z} \right> \hspace*{1 cm}\nonumber \\
\hspace*{2 cm} - Re^{-1} \left< {\bf \nabla v } : {\bf \nabla v } \right>   , \label{energyv} \\
0 = \left< \Phi ' B_x B_z \right> + \left< {\bf j} \cdot ({\bf v} \times {\bf B} ) \right> - R_m^{-1} \left< {\bf j}^2 \right>, \label{energyB}
\end{eqnarray}
where the Poynting flux going out of the control volume has been assumed to be zero (there is no external source or sink, see appendix \ref{appendix2}) and 
where the following global space-time average is introduced for any quantity $f$:
\begin{equation}
<f> = \lim_{\substack{T \rightarrow \infty}} \lim_{\substack{L \rightarrow \infty}} \frac{1}{L^2 T} \int_0^{\substack{T}} \!\!\! \int_{\substack{-1}}^{\substack{1}} \! \int_{\substack{0}}^{\substack{L}} \!\!\! \int_{\substack{0}}^{\substack{L}} f \ dx dy dz dt. \label{average} 
\end{equation}

The work of Lorentz forces $\left< ( {\bf j} \times {\bf B} ) \cdot {\bf v} \right> $ and electromotive work $ \left< {\bf j} \cdot ({\bf v} \times {\bf B} ) \right> $ are equal and opposite, so that the sum of (\ref{energyv}) and (\ref{energyB}) leads to the following energy balance:
\begin{align}
&&- \left< \Phi ' v_x v_z \right> + \left< \Phi ' B_x B_z \right> - Re^{-1} \left<  \Phi ' \frac{ \partial v_x}{\partial z} \right> \hspace*{1 cm} \nonumber \\
&& \hspace*{2 cm}  - Re^{-1} \left< {\bf \nabla v } : {\bf \nabla v } \right> - R_m^{-1} \left< {\bf j}^2 \right> = 0 . \label{balance}
\end{align}

\section{Total energy dissipation}
\label{totenergy}

Let us take $\rho \nu U^2 / H^2$ as a dimensional scale for energy dissipation per cubic metre.  
The average dimensionless energy dissipation due to viscous effects is $\left< {\bf \nabla u } : {\bf \nabla u } \right> $ and the average dimensionless Joule dissipation is $P_m^{-1} \left< {\bf j}^2 \right> $. When ${\bf u}$ is expressed using the decomposition (\ref{decomposition}), the total dissipation ${\cal{D}} = \left< {\bf \nabla u } : {\bf \nabla u } \right> + P_m^{-1} \left< {\bf j}^2 \right> $ becomes
\begin{equation}
{\cal{D}} = \left< {\bf \nabla v } : {\bf \nabla v } \right>  +  \left< \Phi '^2 \right>  + 2  \left< \Phi '  \frac{ \partial v_x}{\partial z} \right> + P_m^{-1} \left< {\bf j}^2 \right> . \label{expr-diss}
\end{equation}
Combining the energy constraint (\ref{balance}), so as to remove the linear term $\left< \Phi '  { \partial v_x} / {\partial z} \right> $, gives the following expression for the dissipation: 
\begin{align}
&& {\cal{D}} =  \left< \Phi '^2 \right> - \left[ \left< {\bf \nabla v } : {\bf \nabla v } \right>  + 2 Re \left< \Phi '  v_x v_z \right>  \right] \hspace*{1 cm} \nonumber \\
&& \hspace*{2 cm} - P_m^{-1} \left[ \left< {\bf j}^2 \right> - 2 R_m \left< \Phi '  B_x B_z \right>  \right]  . \label{diss}
\end{align}
Dissipation is bounded by the background dissipation $\left< \Phi '^2 \right>$ when both conditions $ \left< {\bf \nabla v } : {\bf \nabla v } \right>  + 2 Re \left< \Phi '  v_x v_z \right>  > 0$ and $ \left< {\bf j}^2 \right> - 2 R_m \left< \Phi '  B_x B_z \right> > 0$ are satisfied for all admissible vector fields ${\bf v}$ and ${\bf B}$, {\it i.e.} divergence-free vector fields satisfying the appropriate boundary conditions. Those two conditions are called spectral conditions in the framework of the background method because they can be treated as an eigenvalue problem and this is preceisely the method that will be followed in section \ref{matlabopt}. However, it is also possible to ensure that those conditions are satisfied using other methods and they will be dealt with using functional inequalities in the following sections \ref{velspec} and \ref{magspec}.  \\

\section{Velocity spectral condition}
\label{velspec}

Equation (\ref{diss}) shows that the spectral conditions can be treated independently for velocity and magnetic disturbances. The velocity spectral condition consists in ensuring that $\left< {\bf \nabla v } : {\bf \nabla v } \right>  + 2 Re \left< \Phi '  v_x v_z \right> $ remains positive for all velocity fields ${\bf v}$.  

In order to satisfy the spectral constraint at large Reynolds numbers, it is convenient to choose a background velocity profile which is uniform in most of the fluid, with a linear profile on each side so as to recover correct boundary conditions (see fig. \ref{configuration}). The thickness $\delta$ of these linear parts (virtual boundary layers) is a free parameter. The `production term' $2 Re \left< \Phi '  v_x v_z \right>$ is confined to these regions where it can hopefully be balanced by viscous `dissipation' $\left< {\bf \nabla v } : {\bf \nabla v } \right>$ for a sufficiently small thickness $\delta$. 

This problem is solved as follows (Doering and Gibbons \cite{dg95}, Alexakis et al. \cite{APMD03}). Using the condition ${\bf v} = {\bf 0}$ at the lower wall, velocity at another position $z$ in the fluid can be bounded as follows using Schwartz relationship:
\begin{eqnarray}
v_x &=& \int_{-1+e}^{z} \frac{\partial v_x }{\partial z'} dz' \nonumber \\
& \leq & \sqrt{\int_{-1+e}^{z} dz' \ \int_{-1+e}^{z} \left[ \frac{\partial v_x }{\partial z'} \right]^2  dz'}  \nonumber \\
& \leq & \sqrt{z+1-e} \sqrt{\int_{-1+e}^{1-e} \left[ \frac{\partial v_x }{\partial z'} \right]^2  dz'}. \label{schwartz}
\end{eqnarray}
Combining with the corresponding equation for $v_z$ and using Young relationship leads to:
\begin{eqnarray}
v_x v_z & \leq & \frac{1}{2} (z+1-e) \left[ \int_{-1+e}^{1-e} \left[ \frac{\partial v_x }{\partial z'} \right]^2 + \left[ \frac{\partial v_z }{\partial z'} \right]^2  dz' \right] \nonumber \\
& \leq & \frac{1}{2} (z+1-e)  \int_{-1+e}^{1-e} {\bf \nabla v } : {\bf \nabla v } dz. \label{vxvz}
\end{eqnarray}
Integrating $v_x v_z \Phi '$ over the lower boundary layer yields: 
\begin{equation}
\int_{\substack{-1+e}}^{\substack{-1+e+\delta}} v_x v_z \Phi ' dz  \leq  \frac{\delta}{4} \int_{-1+e}^{1-e} {\bf \nabla v } : {\bf \nabla v } dz \label{vxvzPhi}
\end{equation}
Averaging over the $x$ and $y$ direction and taking into account the upper boundary layer leads to:
\begin{equation}
\left< \Phi '  v_x v_z \right>  \leq \frac{\delta}{2} \left< {\bf \nabla v } : {\bf \nabla v } \right>. \label{nearly}
\end{equation}
  The velocity spectral condition is then satisfied as soon as:
\begin{equation}
\delta \leq Re^{-1}. \label{spectralvitesse}
\end{equation}
In terms of energy dissipation (\ref{diss}), assuming magnetic effects to be absent, this background profile with sufficiently small boundary layers (\ref{spectralvitesse}) provides an upper bound:
\begin{equation}
{\cal{D}} \leq 2 Re \label{hydro-diss}
\end{equation}

\section{Magnetic spectral condition}
\label{magspec}

In the previous section, it has been possible to find a condition on $\delta$ fulfilling the spectral requirement for velocity perturbations. The corresponding task for magnetic perturbations cannot be exactly similar as the property ${\bf v} = {\bf 0}$ on the walls has to be changed into physically sound magnetic boundary conditions. We are considering that the magnetic field inside the fluid and boundary domains should match a potential field outside decaying to zero at infinity (or possibly to a non-zero constant if a non-zero mean electric current density is allowed in the electrically conducting domain). 

We cannot easily find a simple expression for the boundary condition for $B_x$ or $B_z$ in the physical space, however we obtain below a boundary condition for the ($x, y$) average product $B_x B_z$. This will then been used to bound $\left< \Phi '  B_x B_z \right>$ in terms of $\left< {\bf j}^2 \right>$. 

Let us denote the horizontal average over the $(x,y)$ plane with an overbar:
\begin{equation}
\overline{f} (z)=\lim_{\substack{L\rightarrow \infty}} \frac{1}{L^2} \int_0^L \int_0^L f(x,y,z)\,  dxdy. \label{overbar}
\end{equation}
Outside the domain, the curl of the magnetic field is zero. So is the product of its $y$ component with $B_z$ averaged over $x$ and $y$:
\begin{equation}
\overline{B_z (\partial _z B_x - \partial _x B_z)} = 0.\label{magBC}
\end{equation}
 Using integration by part, divergence-free condition for ${\bf B}$, another integration by parts and the $z$ component of ${\bf \nabla}\times{\bf B} = {\bf 0}$ leads to:
\begin{eqnarray}
\overline{B_z (\partial _z B_x - \partial _x B_z)} &=& \partial _z \overline{B_x B_z} - \overline{B_x \partial _z B_z}, \nonumber \\
&=& \partial _z \overline{B_x B_z} + \overline{B_x \partial _y B_y}, \nonumber \\
&=& \partial _z \overline{B_x B_z} - \overline{B_y \partial _y B_x}, \nonumber \\
&=& \partial _z \overline{B_x B_z} + \overline{B_y \partial _x B_y}, \nonumber \\
&=& \partial _z \overline{B_x B_z} \ = \ 0, \label{magBC2} 
\end{eqnarray}
As the product $B_x B_z$ is vanishing at infinite $z$, we conclude that $\overline{B_x B_z}$ is zero at the boundary of the conducting domain, {\it i.e.} when $z=\pm 1$. 
Strictly speaking, this is enough to see that it will be possible to bound $\left< B_xB_z \right> $ in terms of the mean square of the gradient of ${\bf B}$ using Poincar\'e's theorem, and then in terms of the mean joule dissipation. A convenient way of obtaining such a bound is to repeat the calculations leading to (\ref{magBC2}) while retaining a non-zero electric current density for $-1 \leq z \leq 1$. This leads to: 
\begin{equation}
\partial _z \overline{B_x B_z} = \overline{j_y B_z - j_z B_y} = \overline{({\bf j} \times {\bf B})_x}, \label{stresstensor}
\end{equation}
which is the expression of the $x$ component of Lorentz forces in terms of the magnetic stress tensor, averaged on constant $z$ planes. 
The product $\overline{B_x B_z}$ can be evaluated inside the electrically conducting domain using the boundary condition $\overline{B_x B_z}=0$ at $z=\pm 1$. 
\begin{eqnarray}
\overline{B_x B_z} (z) &=& \int_{-1}^z \partial _z \overline{B_x B_z} dz', \nonumber \\
&=& \int_{-1}^z \overline{({\bf j}\times {\bf B})_x}, \label{BxBz}
\end{eqnarray}
which can be bounded as follows:
\begin{equation}
\left| \overline{B_x B_z} (z) \right|  \leq  \sqrt{\int_{-1}^z dz'} \sqrt{\int_{-1}^z \overline{{\bf j}^2} \, \overline{{\bf B}^2} dz'}. \label{BxBzz}
\end{equation}
As shown in appendix \ref{appendix1}, the $x$ and $y$ averaged square magnetic field in the electrically conducting domain is bounded as follows:
\begin{equation}
\overline{  {\bf B}^2 } (z) \leq 2 \left< {\bf j}^2 \right> . \label{boundB2}
\end{equation}
Hence, equation (\ref{BxBzz}) can be written:
\begin{equation}
\left| \overline{B_x B_z} (z) \right| \leq \sqrt{2} \sqrt{z+1} \left< {\bf j}^2 \right>. \label{BxBzf}
\end{equation}
The magnetic boundary condition is rather loose compared to the velocity boundary condition which leads here to a bound proportional to the square-root of $z+1$ for the product $\overline{B_x B_z}$ rather than a linear dependence for the velocity product $\overline{v_x v_z}$.  

\begin{figure}[ht]
\begin{center}
\includegraphics[width=8.5 cm, keepaspectratio]{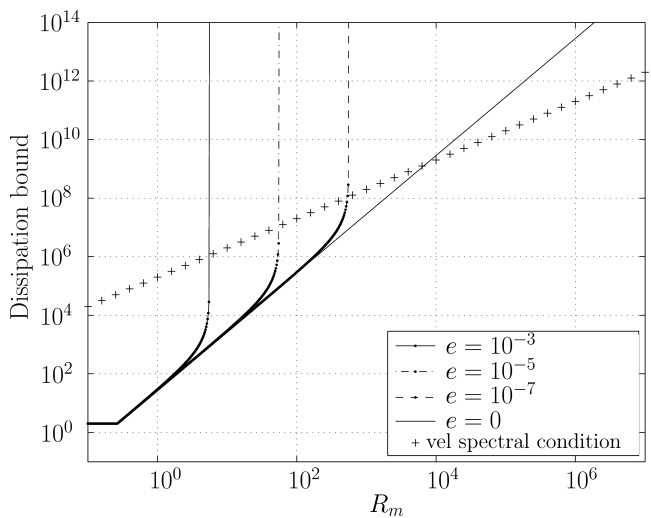}
\caption{Bound on dissipation obtained from equations (\ref{hydro-diss}) for the velocity spectral condition (`plus' signs), and (\ref{magneticspectralb}) for the magnetic spectral condition, for different wall conductances $e=10^{-3}$, $10^{-5}$, $10^{-7}$ and $0$, plotted here for $P_m = 10^{-5}$}
\label{dissanalfig}
\end{center}
\end{figure}

Using equation (\ref{BxBzf}) it is now possible to bound $\left< B_x B_z \Phi ' \right>$:
\begin{eqnarray}
\left| \left< B_x B_z \Phi '
\right> \right| & \leq & \frac{1}{\delta} \int_{-1+e}^{-1+e+\delta} \left| \overline{B_x B_z} (z) \right| dz \nonumber \\
&&\hspace*{1.2 cm} + \frac{1}{\delta} \int_{1-e-\delta}^{1-e} \left| \overline{B_x B_z} (z) \right| dz , \nonumber \\
& \leq & \frac{2 \sqrt{2}}{\delta} \left< {\bf j}^2 \right> \int_e^{e+\delta} \sqrt{u} d u.  \nonumber \\
& \leq & \frac{4 \sqrt{2}}{3 \delta} \left< {\bf j}^2 \right> \left[ (e+\delta )^{3/2} -e^{3/2} \right]\label{BxBzPhip}
\end{eqnarray}
The magnetic spectral condition $ \left< {\bf j}^2 \right> - 2 R_m \left< \Phi '  B_x B_z \right> \geq 0 $ is satisfied as soon as:
\begin{equation}
\frac{8 \sqrt{2} \, R_m}{3 \delta } \left[ (e+\delta )^{3/2} -e^{3/2} \right] \leq 1. \label{magneticspectral}
\end{equation}
This relationship is then used to express $\delta$ is terms of viscous dissipation for the background flow, hence to determine an upper bound for global dissipation. This bound is plotted on Fig. \ref{dissanalfig} for various values of the wall thickness $e$. The limiting case of vanishing wall thickness (or rather of an electrically insulating wall) can be easily derived from (\ref{magneticspectral}) and corresponds to:
\begin{equation}
\frac{8 \sqrt{2} \, R_m \, \delta ^{1/2}}{3 } \leq 1, \label{magneticspectralb}
\end{equation}
and since the dissipation of the background profile is $2/\delta$, we have:
\begin{equation}
{\cal{D}} \leq \frac{256 \, R_m^2 }{9}. \label{magneticspectralc}
\end{equation}

One must take the higher upper bound derived from the velocity and magnetic spectral conditions in Fig. \ref{dissanalfig} as both conditions must be satisfied. It is straightforward from equations (\ref{hydro-diss}) and (\ref{magneticspectral}) that the magnetic spectral condition will take over the velocity spectral condition when the magnetic Reynolds number is exceeding $9  / ( 128 \, P_m ) \simeq 0.07 \, P_m^{-1}$.

It can also be derived from (\ref{magneticspectral}) that the critical magnetic Reynolds number at which our upper bound estimate diverges towards infinity is $R_m = 1 / (4 \sqrt{2 e})$. 

\section{Numerical determination of upper bounds}
\label{matlabopt}

The velocity and magnetic spectral conditions appearing in (\ref{diss}) are tackled numerically in this section. The reason for this is not to achieve a better ({\it i.e.} lower) upper bound, although this is the case. The main reason is to check that our analysis in sections \ref{velspec} and \ref{magspec} provides results that are of the correct magnitude and not gross overestimates of dissipation upper bounds. For a given background profile, the numerical analysis provides Reynolds and magnetic Reynolds numbers such that the spectral conditions are satisfied for all admissible velocity and magnetic fluctuations, and this cannot be improved as one particular velocity disturbance and one particular magnetic disturbance make the spectral conditions just zero. 

\begin{figure}[ht]
\begin{center}
\includegraphics[width=8.5 cm, keepaspectratio]{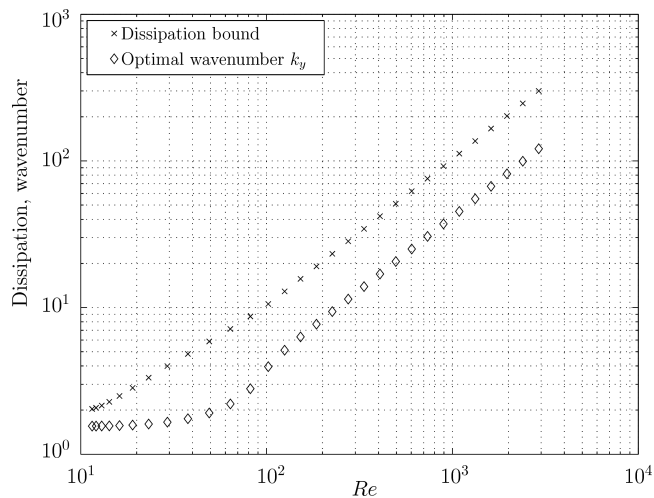}
\caption{Numerical upper bound for energy dissipation from the velocity spectral condition and associated spanwise wave-number $k_y$ of marginal perturbation}
\label{diss-vel}
\end{center}
\end{figure}

The method is sketched here: the spectral condition can be treated as a problem of energy stability for the background flow \cite{dg95}. This is changed into an eigenvalue problem and the game consists in finding a Reynold or magnetic Reynolds number such that the maximum eigenvalue is zero. The velocity and magnetic fluctuations are Fourier transformed in the $x$ and $y$ directions, while they are expanded using Chebyshev collocation polynomials in the $z$ direction. The optimal magnetic disturbance is always found in the limit of zero wave numbers $k_x$ and $k_y$, whereas the optimal velocity disturbance is found for $k_x=0$ and $k_y$ finite and increasing with the Reynolds number (see Fig. (\ref{diss-vel})). 

\begin{figure}[ht]
\begin{center}
\includegraphics[width=8.5 cm, keepaspectratio]{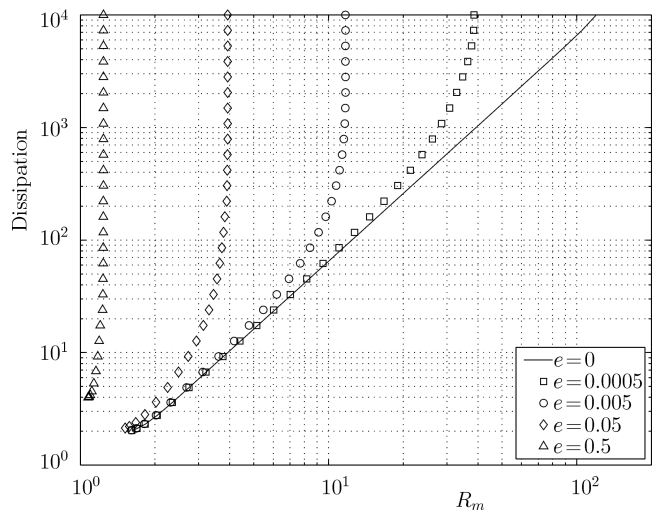}
\caption{Numerical magnetic upper bound for energy dissipation for different wall thicknesses}
\label{diss251}
\end{center}
\end{figure}

For this numerical analysis, the background function is no longer piecewise linear. In order to avoid numerical difficulties due to discontinuities in the velocity gradient, a hyperbolic sine function class is chosen $\Phi = \sinh (z \delta)$. In the limit of large dissipation, the background velocity is zero everywhere except in thin boundary layers where the velocity profile is an exponential function. Similarly to what has been done in sections \ref{velspec} and \ref{magspec}, the background functions contain a single free parameter, {\it i.e}, the typical thickness $\delta$ of the boundary layers. Due to the new profile of the background function, the associated dissipation takes a different form that can be derived analytically: 
\begin{equation}
\frac{1 - e + \sinh \left( 2 \frac{1-e}{ \delta} \right) }{ 2 \delta \left[ \sinh \left( \frac{1-e}{\delta} \right) \right]^2 }. \label{dissSinh}
\end{equation}
For a given choice of $\delta$, hence dissipation, a lower bound of Reynolds and magnetic Reynolds number is sought. This is repeated for various values of $\delta$ and the resulting curve can be read as an upper bound of dissipation for each value of Reynolds or magnetic Reynolds number (see Fig. \ref{diss-vel} and Fig. \ref{diss251} respectively). Both spectral conditions can be plotted on the same figure once a magnetic Prandtl number is specified. 

It is clear that this numerical procedure leads to improved upper bounds -- by a factor 10 approximately -- compared to those obtained analytically, while similar trends are obtained.

\section{Discussion}
\label{discussion}

For the simple Couette flow configuration under investigation, and for electrically insulating boundaries, it has been found that a classical 'hydrodynamical' upper bound for dissipation holds when $R_m$ is smaller than around $P_m^{-1}$: the dimensional upper bound of dissipation per unit mass is of order $U^3/H$. For larger magnetic Reynolds numbers though, another scaling is followed and the bound on dissipation becomes proportional to $\mu ^2 \sigma ^2 \nu \, U^4$. This new bound is higher than the 'hydrodynamical' bound and is independent of the half-distance $H$ between the plates. 

When the boundary plates have the same electrical conductivity as the fluid and a finite thickness $e$, the resulting upper bound is changed dramatically. The upper bound corresponding to the magnetic part of the spectral condition increases suddenly and diverges to infinity for a value of the magnetic Reynold number depending on the thickness of the walls. This critical magnetic Reynolds number scales as $e^{-1/2}$. This value can be reached much before the change of regime between 'hydrodynamical' and (insulating) 'magnetic' branches of upper bound estimates ($R_m \sim P_m^{-1}$). 

It should not be too surprising that dissipation can be unbounded at some finite magnetic Reynolds number. Such a situation can be easily simulated with a Bullard dynamo model. This is a solid-rotation dynamo with no possible feedback of the magnetic field on the structure of the flow. Hence magnetic energy (and dissipation) grows without limit above the threshold of linear instability. As soon as electrical conductivity is restricted to the fluid ($e=0$), this behaviour is no longer possible. For any finite magnetic Reynolds number, energy dissipation is certainly finite. Yet the scaling law for this dissipation is distinctly above the 'hydrodynamical' {\it a priori} estimate.  

Whether one is inclined to believe that upper bounds of dissipation are generally relevant to turbulence or not, there is one firm and strong result from the present work which is the extended validity of the Kolmogorov scaling up to $R_m \sim P_m^{-1}$. It might have been a reasonable guess to predict a distinct increase of the dissipation as soon as $R_m$ exceeds a critical threshold of order unity. This analysis shows that this is not possible with electrically insulating boundaries. This certainly has implications on the design of future Couette-type dynamo experimental setups, depending on whether a small or large dissipation is sought. Although there is no mathematical statement on how dissipation is divided between mechanical and electrical phenomena, it might be a better choice to have electrically conducting boundaries, should one want to increase the ratio of Joule to viscous dissipation.   

It is appropriate to discuss here how energy dissipation was affected by dynamo action in those three experimental setups where a self-sustained liquid dynamo was observed. They are the Riga (Latvia) \cite{rigaDYN}, Karlsruhe (Germany) \cite{karlsruheDYN} and Cadarache (France) \cite{cadarache} dynamos. From these references, a table can be shown (table \ref{tablediss}), where the estimated extra dissipation when dynamo action is present is given (in percentage) for a certain magnetic Reynolds number (expressed in percentage above the threshold). the Karlsruhe dynamo experiment can be clearly distinguished by the large amount of extra dissipation compared to the other two. This experiment is also one where differential velocity is forced deep inside the setup. For the Riga and Cadarache dynamos, a turbine or propellers are set in rotation near the outer boundary of the setup. There can only be a gross comparison to the pure Couette flow configuration presented here, but the case of thick electrically conducting boundaries is also putting shear velocities well inside the electrically conducting domain and results in a huge increase of the dissipation bound. Conversely, it cannot be ruled out that Riga and Cadarache dynamos are still obeying the same classical 'hydrodynamical' law of dissipation, similarly to the Couette flow with electrically insulating walls. 
\begin{table}
\caption{Extra dissipation due to dynamo action slightly above threshold in experimental dynamos}
\label{tablediss}
\begin{center}
\begin{tabular}{|l|c|c|c|}
\hline
 & Riga \cite{rigaDYN} & Karlsruhe \cite{karlsruheDYN} & Cadarache \cite{cadarache} \\
\hline
\begin{minipage}{1.8 cm}{\% above critical $R_m$}\end{minipage} & \makebox[1.8 cm]{6 -- 7}        & \makebox[2 cm]{6.7}   & \makebox[2 cm]{30}  \\
\hline
\begin{minipage}{1.8 cm}{\% extra dissipation}\end{minipage}    & 10            & 42    & 15 -- 20 \\
\hline
\end{tabular}
\end{center}
\end{table}
The authors of the dynamo experiments themselves specify that 
the extra dissipation is obtained by difference of the actual dissipation and the extrapolated curve of dissipation below threshold. Even though it has been used to estimate Ohmic dissipation by some authors, it should be stressed here that there is no theoretical reason why this extra dissipation should be the Ohmic contribution to dissipation. For instance, in the present work, we use the fact that there is an exact cancellation between the work of Lorentz forces and the electromotive work, but we have no access to either of these terms.

We have concentrated mainly on the low magnetic Prandtl number case, because our primary interest lies in the Earth's core dynamics and in the relevant liquid metal experiments. Nevertheless the results we have obtained are equally valid for large magnetic Prandtl numbers. It is however anticipated that our dissipation bound will be a gross overestimate at low values of the hydrodynamic Reynolds number. In that case, the flow is the simple uniform shear flow between the moving walls and such a flow is not going to sustain dynamo action easily. This perspective should not be ruled out as strong Lorentz forces might change the flow in such a way that it could drive a dynamo, however this transition -- if it exists -- will be severely subcritical.  


Saturation can somehow be addressed within this work. 
The total energy dissipation is an upper bound for Joule dissipation, which can be used through equation (\ref{A1}) to obtain an upper bound of the magnetic energy in a saturated regime of dynamo action. This bound cannot give realistic estimates near dynamo threshold, as dissipation is still dominated by the viscous contribution. At large magnetic Reynolds numbers, this bound is much larger than equipartition between kinetic and magnetic energy.

In the derivation of velocity and magnetic spectral conditions, there are strong similarities. The only difference is due to boundary conditions for velocity disturbances and magnetic field disturbances. This mere difference in the boundary conditions leads to fundamentally different upper bounds for dissipation. Magnetic boundary conditions are sometimes not well treated in general textbooks (see, for instance, the energy stability of hydromagnetic flows in the otherwise excellent book by Joseph \cite{joseph2}, volume II). It is confirmed again in the present work that one should pay great attention to the relevant physical boundary conditions.  

%
%

%
%

\section{Perspectives}
\label{perspectives}

It is certainly useful to derive such upper bounds for dissipation in the field of magnetofluid dynamics, and more particularly regarding dynamos. When complex situations are considered (magnetic field, global rotation, ...) there is no reliable heuristic approach that can provide good estimates for dissipation. Ordinary hydrodynamic turbulence can be roughly tackled by Kolmogorov's theory but it is not safe to extend it to other types of turbulence. When the 'Doering-Constantin-Hopf' background profile method can provided an upper bound, this is a solid reliable result. The bound can be rather constraining for instance in the case of a Couette flow with electrically insulating boundaries, at low $P_m$: it is not an obvious result that dissipation  must remain similar to hydrodynamic dissipation up until $R_m \sim P_m^{-1}$. There are actually a lot of other configurations for which the method can be applied.  

Upper bounds for dissipation can be calculated for experimental setups involving liquid metal flows. In Grenoble, LGIT laboratory, we have the DTS setup involving 40 liters of sodium in an imposed magnetic field \cite{dts-super,NABCGJS07,SABCGJN07}. Upper bounds for dissipation would provide some information about the importance of size scaling of the experimental setup. This is crucial when planning a larger setup that might sustain dynamo action. 

Thermal convection can also be taken into account to derive dissipation bounds.
This is the driving force for the flow of liquid iron in the Earth's core. Upper bounds of dissipation may prove useful in terms of the thermal budget of Earth  throughout its history. Although the configuration of the core of the Earth is not similar to the simple Couette flow considered here, it could be very relevant to examine the role of a thin electrically conducting layer at the bottom of the mantle, as it could make a large change in the upper bound for dissipation. \\

\begin{acknowledgments} 
Thanks are due to Alexia Gorecki and Maylis Landeau who have been working on student projects on closely related topics and provided stimulating discussions. I have benefited from a discussion with Christophe Gissinger on the role of the Poynting flux and with Alexandros Alexakis who pointed out to me a mistake in the original proof regarding the analytical bound for the magnetic spectral constraint. I am also grateful to the members of the Geodynamo team (LGIT lab in Grenoble) for their comments on this work.  
\end{acknowledgments}

\appendix

\section{Magnetic boundary conditions and associated Poynting flux}
\label{appendix2}

Outside our electrically conducting domain $-H < z < H$ ({\it i.e.} the fluid layer and electrially conducting part of the plates), it is assumed that there exist no other electrically conducting domains, and no regions with magnetic properties: let us just think of it as empty space. Coherent with the classical magnetostatic approximation used in the electrically conducting domain to derive the induction equation (\ref{induction}), the displacement current is also neglected outside, so that the magnetic field satisfies the following Maxwell equations:\begin{equation}
{\bf \nabla }\cdot {\bf B} = 0 \hspace{1.5 cm} {\bf \nabla }\times {\bf B} = {\bf 0},  \label{Bbc}
\end{equation}
from which it follows that the magnetic field is harmonic. The magnetic field can be written as a Fourier integral:
\begin{equation}
{\bf B} = \int {  \int { \tilde{\bf B} (k_x, k_y, z, t)\  e^{(i k_x x + i k_y y)} dk_x } \, dk_y },   \label{fourierB}
\end{equation}
and the harmonicity of ${\bf B}$ implies that each component ($k_x$ and $k_y$ fixed) must satisfy:
\begin{equation}
\frac{d^2 \tilde{\bf B}}{d z^2} = - k^2 \tilde{\bf B}, \label{expon}
\end{equation}
where $k^2=k_x^2+k_y^2$. 
As a consequence $\tilde{\bf B}$ must decay exponentially away from the fluid layer on a typical distance $1/k$. There is one exception when there is a significant magnetic contribution near $k=0$. However, we shall exclude that possibility on physical grounds: in that case, the exponential decay length increases without limit and there would be an infinite amount of magnetic energy stored oustside the fluid. That would take an infinitely long time to settle and is thus incoherent with our assumption of the existence of a stationary flow solution. 

Faraday's equation (another equation from the set of Maxwell's equations) is used to determine the electric field:
\begin{equation}
{\bf \nabla}\times {\bf E} = - \frac{\partial {\bf B}}{\partial t},  \label{electric}
\end{equation}
showing that it must also decay exponentially. Hence, the Poynting flux density ${\bf E}\times{\bf B}/\mu$ decays exponentially and is thus as close to zero as one wishes some distance away from the fluid layer. From the assumption of statistical stationarity, there can be no accumulation of magnetic energy, hence this Poynting flux ${\bf E} \times{\bf B}/\mu$ must vanish when averaged along $x$ and $y$, at any position $z$ outside the electrically conducting domain. This reasoning is valid for each wavenumber and the total Poynting flux is the sum of individual wavenumber components, hence this proves that the averaged Poynting flux must be zero at any position $z$ and in particular at the boundary $z=H$ and $z=-H$ (or $z=\pm 1$ in dimensionless coordinates). Let us insist that this result is true when the Poynting flux is averaged on a plane of constant $z$ (not pointwise) and when it is also averaged in time, assuming stationary turbulence. 

It should be noted that the above reasoning fails in two circumstances. One is when there are other electrically conducting domains. In that case, magnetic energy can be dissipated or generated in each domain and there can be a net exchange of energy between the fluid layer and the other domains. The other possibility is that energy is radiated away from the fluid layer and goes away infinitely far. This is only possible when the displacement current is taken into account in Ampere's equation (\ref{Bbc}). Then a significant fraction of energy can be radiated away only when the timescale $\tau$ and lengthscale $l$ of the magnetic field are such that $l/\tau$ is comparable to the speed of light \cite{houches07}. This is a necessary condition for our system to work as an antena, and we shall not consider this limiting case.  

\section{Bound on magnetic energy for a given amount of Joule dissipation}
\label{appendix1}

We show here that magnetic energy in the conducting domain is bounded pointwise by the integral Joule dissipation. In our dimensionless terms, the following relationship hods for any value of $z$:
\begin{equation}
\overline{ {\bf B}^2 } (z) \leq 2 \left< {\bf j}^2 \right> . \label{A1}
\end{equation}
The magnetic field ${\bf B}$ is decomposed in polo\" \i dal-toro\" \i dal contributions: 
\begin{equation}
{\bf B} = {\bf \nabla } q \times {\bf e}_z + {\bf \nabla } \times \left(  {\bf \nabla } p \times {\bf e}_z \right) + b_{0x} (z) {\bf e}_x + b_{0y} (z) {\bf e}_y
\end{equation}
The assumption of spatial statistical invariance along $x$ and $y$ directions allows us to write the polo\" \i dal and toro\" \i dal contributions of ${\bf B}$ in terms of a Fourier integral
\begin{equation}
\left[ p,\, q \right] =  \int { \! \!  \int { \left[ P,\, Q \right] (k_x, k_y, z, t)\  e^{(i k_x x + i k_y y)} dk_x } \, dk_y } \hspace{1 cm}
\end{equation}
Each Fourier component can be considered individually, as magnetic energy and ohmic dissipation will be just the sum of energy and dissipation of each contribution. It is enough to prove (\ref{A1}) for all contributions. The contribution $ b_{0x} (z) {\bf e}_x + b_{0y} (z) {\bf e}_y $ will be dealt with at the end of this appendix. Polo\" \i dal and toro\" \i dal contributions lead to 
\begin{equation}
{\bf B} = \left| \begin{array}{l} -\partial _y q - \partial ^2 _{xz} p \\ \partial _x q - \partial ^2 _{yz} p \\ {\nabla}^2_S p \end{array} \right.
\end{equation}
\begin{equation}
{\bf j} = {\bf \nabla} \times {\bf B} = \left| \begin{array}{l} \partial _y {\nabla}^2 p - \partial ^2 _{xz} q \\ - \partial _x {\nabla}^2 p - \partial ^2 _{yz} q \\ {\nabla}^2_S q \end{array} \right.
\end{equation}
Hence, $x$ and $y$ averaged magnetic energy and ohmic dissipation can be written
\begin{equation}
\overline{{\bf B}^2} (z) = \frac{k^2}{2} \left| Q \right| ^2 + \frac{k^2}{2} \left| \partial _z P  \right| ^2 + \frac{k^4}{2} \left| P \right| ^2 \label{magA}
\end{equation}
\begin{equation}
\overline{{\bf j}^2} (z) = \frac{k^2}{2} \left| - k^2 P +  \partial ^2_z P  \right| ^2 + \frac{k^2}{2} \left| \partial _z Q \right| ^2  + \frac{k^4}{2} \left| Q \right| ^2
\end{equation}
Integration by parts leads to the following expression for the global dissipation using the condition that the polo\" \i dal and toro\" \i dal components must decay far away from the fluid layer
\begin{eqnarray}
\left< {\bf j}^2 \right> &=& \int_{-1}^1 \overline{{\bf j}^2} dz = \int_{-\infty}^\infty \overline{{\bf j}^2} dz \nonumber  \\
&=& \int_{-\infty}^\infty \frac{k^2}{2} \left|  \partial^2 _{zz} P \right| ^2 + k^4 \left| \partial _z P \right| ^2 + \frac{k^6}{2} \left| P \right| ^2 dz \nonumber \\
&&  +  \int_{-\infty}^\infty \frac{k^2}{2} \left| \partial _z Q \right| ^2 + \frac{k^4}{2} \left| Q \right| ^2 dz \label{dissA}
\end{eqnarray}
The toro\" \i dal scalar function $Q$ must vanish at $z=\pm1$. Hence:
\begin{equation}
Q(z) = \int_{-1}^z \partial _z Q dz 
\end{equation}
Hence by Cauchy-Schwartz
\begin{eqnarray}
\left| Q \right|^2 &\leq &  \int_{-1}^z dz \int_{-1}^z \left| \partial _z Q \right| ^2 dz  \nonumber  \\
&\leq &  2 \int_{-1}^1 \left| \partial _z Q \right| ^2 dz \label{torA}
\end{eqnarray}
The treatment of the polo\" \i dal is slightly more involved as the boundary conditions available are less straightforward: $\partial _z P \pm P = 0$ at $z=\pm 1$. They can be used as follows
\begin{eqnarray}
2 k P &=& \left( k P - \partial _z P \right) + \left( k P + \partial _z P \right) \nonumber \\
&=& \int_{-1}^z \! k \partial _z P - \partial ^2_{zz} P dz' - \int_{z}^1 \!  k \partial _z P + \partial ^2_{zz} P dz' \hspace*{10 mm}
\end{eqnarray}
from which the modulus of $P$ can be bounded using the triangle and Cauchy-Schwartz inequalities
\begin{eqnarray}
2 k \left| P \right|  & \leq & \int_{-1}^1 k \left| \partial _z P \right| + \left| \partial ^2_{zz} P \right| dz' \nonumber \\
& \leq & \sqrt{ 2 \int _{-1}^1 \! k^2 \left| \partial _z P \right| ^2 dz'  } + \sqrt{ 2 \int _{-1}^1  \! \left| \partial ^2 _{zz} P \right| ^2 dz'  } \hspace{1 cm}
\end{eqnarray}
Hence
\begin{equation}
k^2 \left| P \right| ^2 \leq k^2 \int _{-1}^1 \left| \partial _z P \right| ^2 dz' +  \int _{-1}^1 \left| \partial ^2 _{zz} P \right| ^2 dz' \label{polA1}
\end{equation}
A similar treatment is made on $\partial _z P$
\begin{equation}
2 \partial _z P = - \left( k P - \partial _z P \right) + \left( k P + \partial _z P \right) 
\end{equation}
leading to a similar result
\begin{equation}
\left| \partial _z P \right| ^2 \leq k^2 \int _{-1}^1 \left| \partial _z P \right| ^2 dz' +  \int _{-1}^1 \left| \partial ^2 _{zz} P \right| ^2 dz' \label{polA2}
\end{equation}
All three terms in (\ref{magA}) can be bounded using (\ref{torA}), (\ref{polA1}) and (\ref{polA2}) and then compared to the contributions of the ohmic dissipation in (\ref{dissA}), leading to the expected result:
\begin{equation}
\overline{ {\bf B}^2 } (z) \leq 2 \left< {\bf j}^2 \right>
\end{equation}
Importantly, this result is independent of $k$, so that it will equally apply to a sum of contributions of different wavenumbers $k_x$ and $k_y$, hence proving the expected result for the fields ${\bf B}$ and ${\bf j}$.

The horizontally independent contributions $b_{0x} (z) {\bf e}_x + b_{0y} (z) {\bf e}_y$ must now be considered. Their associated ohmic dissipation can be written
\begin{equation}
\left< {\bf j}^2 \right> = \int_{-1}^1 \left| \frac{ d \, b_{0x}}{d\, z} \right| ^2 + \left| \frac{ d \, b_{0y}}{d\, z} \right| ^2 dz \label{unifdissA}
\end{equation}
If the global electrical current flux in the $x$ and $y$ direction is not zero, then this is an unphysical situation as magnetic energy in the space above and below the fluid would be infinite, and such a situation would take an infinitely long time to be established. If it is zero, then $b_{0x}$ and $b_{0y}$ vanish on $z=\pm 1$, and one can derive a bound for $\overline{\left| {\bf B} \right| ^2 }$ using Cauchy-Schwartz inequality
\begin{eqnarray}
\left| b_{0x} \right| ^2 (z) &=& \left| \int_{-1}^z \frac{d \, b_{0x} }{ d \, z} dz  \right| ^2 
 \leq  2 \int_{-1}^1 \left|  \frac{d\, b_{0x} }{ d\, z} \right| ^2 dz \hspace{1 cm} \label{unifBA}
\end{eqnarray}
A similar inequality holds for $b_{0y} $, so comparing (\ref{unifdissA}) and (\ref{unifBA}) is also compatible with the inequality to prove.

\bibliography{/home/talbouss/biblio/bib}

\end{document}